\documentclass[twocolumn,aps,showpacs,prb,tightenlines,amsmath,amssymb]{revtex4}
\usepackage{graphicx}
\usepackage{amssymb}
\usepackage{dcolumn}
\usepackage{amsmath}
\usepackage{bm}
\usepackage{colordvi}
\newcommand{\bgreek}[1]{\mbox{\boldmath$#1$\unboldmath}}

\begin{document}              

\title{Singlet-triplet relaxation in SiGe/Si/SiGe double
  quantum dots}
\author{L. Wang}
\affiliation{Hefei National Laboratory for Physical Sciences at
Microscale and Department of Physics,
University of Science and Technology of China, Hefei,
Anhui, 230026, China}
\author{M. W. Wu}
\thanks{Author to  whom correspondence should be addressed}
\email{mwwu@ustc.edu.cn.}
\affiliation{Hefei National Laboratory for Physical Sciences at
Microscale and Department of Physics, University of Science and Technology of China, Hefei,
Anhui, 230026, China}

\date{\today}

\begin{abstract}
We study the singlet-triplet relaxation due to the
 spin-orbit coupling assisted by the electron-phonon scattering in
two-electron SiGe/Si/SiGe double quantum dots in the presence of an
external magnetic field in either Faraday or Voigt configuration.
By explicitly including
the electron-electron Coulomb interaction and the valley splitting
induced by the interface scattering, we employ the
exact-diagonalization method to obtain the energy spectra and the
eigenstates. Then we calculate the relaxation rates with the Fermi
golden rule.  We find
that the transition rates can be
effectively tuned by varying the external magnetic field and the interdot
distance. Especially in the vicinity of the
anticrossing point, the transition rates show intriguing
features. We also investigate the electric-field dependence of the
transition rates, and find that the transition rates are almost
independent of the electric field. This is of great importance in the
spin manipulation since the lifetime remains almost the same during
the change of the qubit configuration from $(1,1)$ to $(2,0)$ by
the electric field.
\end{abstract}


\maketitle
\section{INTRODUCTION}
Spin-based qubits utilizing semiconductor quantum dots
(QDs) are believed to be the prospective candidate for quantum information
processing.\cite{loss,hanson,wiel,reimann} Recently,
silicon QDs have attracted much attention due to their
outstanding spin-related
properties.\cite{culcer,li,prada,pan,liu,shaji,culcer2,xiao,wang,raith,das,podd,lai,wild,
  reboredo,shin,lim,thalakulam,hu2,simmons} 
Specifically, the hyperfine interaction can be reduced by isotopic
purification.\cite{taylor} The Dresselhaus spin-orbit
coupling (SOC)\cite{dresselhaus} is absent thanks to the
bulk-inversion symmetry and the SOC induced by the interface-inversion
asymmetry (IIA) is rather weak.\cite{vervoort,vervoort2,nestoklon}
Moreover, the electron-phonon interaction, which plays an important role
in spin relaxation, is much weaker than that in III-V semiconductor QDs since
there is no piezoelectric interaction in silicon.\cite{li} All these
special properties together lead to a long decoherence time in silicon
QDs, which is of great help in the process of coherent manipulation
and information storage. Furthermore, the physics in silicon is
considerably rich owing to the presence of the valley degrees of
freedom. Silicon has sixfold degenerate conduction band minima, which
can be splitted by either strain or confinement in quantum wells into two
parts: a fourfold-degenerate subspace of higher energy and a
twofold one of lower energy. The twofold
degeneracy can be further lifted by a valley-splitting energy due to
the interface scattering. The valley-splitting energy has a strong
dependence on the confinement length of the
structure.\cite{boykin,friesen}

Nowadays, spin qubits in silicon single and double QDs have been
actively investigated.\cite{culcer,li,prada,pan,liu,shaji,culcer2,xiao,wang,raith,das,podd,lai,wild,reboredo,shin,
lim,thalakulam,hu2,simmons}
In silicon single QDs, we have studied the singlet-triplet (ST) relaxation by explicitly
including the electron-electron Coulomb interaction and the
multivalley effect. Our results in the Voigt configuration agree quite well with the
recent experiment by Xiao {\em et al.}.\cite{xiao} Silicon double QDs,
which have been proven very useful in exploiting the spin
Coulomb blockade,\cite{taylor2} have also attracted much
attention. Recently, Raith {\em et al.}\cite{raith}
studied the magnetic-field and interdot-distance dependences of spin
relaxation in single-electron Si/SiGe double QDs. Li
{\em et al.}\cite{li} calculated the exchange coupling between the
unpolarized triplet and the singlet on the basis of a large
valley splitting. Culcer {\em et al.}\cite{culcer} investigated the multivalley
effect on the feasibility of initialization and manipulation of ST
qubits, showing that the valley degree of freedom makes the physics of Si QDs
quite different from that of single-valley ones. In their work, they
analyzed the spectrum with the lowest few basis functions. As will be shown
in this paper, these lowest basis functions are enough for the
convergence of the energy spectrum under investigation,
 but are inadequate to study the ST relaxation
time, similar to the situation of III-V semiconductor-based QDs.\cite{shen}
 The relaxation rates calculated with
these lowest basis functions and the convergent 
ones differ by about four orders of magnitude.
Therefore, it is necessary to employ
the exact-diagonalization approach with a large number of basis
functions in order to have the correct ST relaxation rates.
 Moreover, the electron-electron Coulomb 
interaction, which is crucial to the energy spectra and the
wavefunctions of the singlet and triplet  eigenstates,
 was not explicitly calculated in the literature but rather given
as a Hubbard parameter.\cite{li,culcer,das}

In this work, we calculate the two-electron ST relaxation 
in SiGe/Si/SiGe double QDs by explicitly including the
Coulomb interaction, the valley degree of freedom as well as the
source of the ST relaxation, i.e., the SOCs.\cite{rashba,nestoklon} 
 We employ the exact-diagonalization method to obtain
 the energy spectrum and the
Fermi golden rule to calculate the spin relaxation rates.\cite{cheng,shen} 
Without losing generality, we focus on a large valley splitting case where the lowest
singlet and three triplet states are all constructed by the lowest
valley eigenstate. We investigate the double QD
 system with either a perpendicular magnetic field (the Faraday configuration)
or a parallel one (the Voigt configuration). We find that the energy
levels of the lowest singlet and three triplet states have a strong
dependence on the external magnetic field and the 
interdot distance. The perpendicular magnetic field affects the energy
levels mainly by the orbital effect and the Zeeman splitting while the
parallel magnetic field only influence the energy levels via the
Zeeman splitting due to the
strong confinement along the growth direction. The interdot distance
has a strong influence on the Coulomb interaction and the orbital
energy. Besides, we also find that
the transition rates of the channels among these four
levels can be markedly modulated by the external magnetic field and the
interdot distance. Moreover, from the dependences of the energy
spectrum on the magnetic field and the interdot distance, we observe the
anticrossing points between the
singlet and one of the triplets. In the vicinity of the anticrossing point, the
transition rates of the channels relevant to these two states present
either a peak or a valley. Furthermore, we also study the effect of
the electric field on the energy levels and the transition rates. With
the increase of the electric field, we find that the configurations of the
lowest four levels change from $(1,1)$ to $(2,0)$, where $(n,m)$ indicates the
numbers of occupancy of the left and right dots. Very different from the
magnetic-field and interdot-distance dependences, we find that the
transition rates are almost independent of the electric field. This property
is of great importance in the spin manipulation, since the
lifetime remains almost unchanged during the variation of qubit configurations.

This paper is organized as follows. We set up the model and lay
out the formalism in Sec.~II. In Sec.~III, we employ the 
exact-diagonalization method
to calculate the energy spectrum and the Fermi golden rule to obtain the ST
relaxation rates. We investigate the magnetic-field (in both
the Faraday and Voigt configurations), the interdot-distance and
the electric-field dependences of the transition rates. The features
of the transition rates in the vicinity of the anticrossing points are also
emphasized. Finally, we summarize in Sec.~IV.

\section{MODEL AND FORMALISM}
\begin{figure}[bth]
\includegraphics[width=6cm]{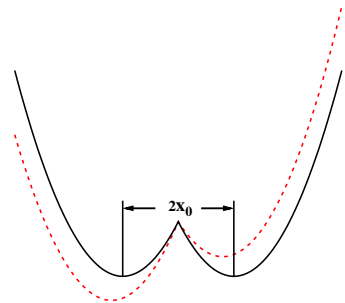}
\caption{(Color online) Schematic of the double QDs. The black solid
  curve stands for the case without applied electric field. 
The red dotted curve represents the case with an
  electric field along the $x$ direction. $2x_0$ is the
  interdot distance.}  
\label{fig1}
\end{figure}

In our model, we choose the lateral confinement potential as
$V_c(x,y)=\frac{1}{2}m_t\omega_0^2\{{\rm min}[(x-x_0)^2,(x+x_0)^2]+y^2\}$ with
$m_t$ and $\omega_0$ representing the in-plane effective
 mass and the confining potential frequency.\cite{fock,darwin} A
 schematic of the double QDs is shown in Fig.~\ref{fig1}. The two
 dots are located at ${\bf R}_{R,L}=(\pm x_0,0,0)$ with $2x_0$ being the
 interdot distance. Here $R$ and $L$ denote right and left,
 respectively. Along the growth direction $[001]$,
 $V_z(z)$ is applied within the infinite-depth well potential
 approximation. The single-electron Hamiltonian with magnetic field
 ${\bf B}=B_\perp\hat{\bf z}+B_\|\hat{\bf x}$ can be written as 
 \begin{eqnarray}
   H_{\rm e}&=&\frac{{P_x}^2+{P_y}^2}{2m_t}+\frac{{P_z}^2}{2m_z}+V({\bf
     r})+H_{\rm so}({\bf P})\nonumber
   \\ &&\mbox{}+H_{\rm Z}+H_{\rm E}+H_{\rm v},
   \label{eq1}
 \end{eqnarray}
with $m_z$ representing the effective mass along the $z$
direction. $V({\bf r})=V_c+V_z$ and ${\bf P}={\bf p}+(e/c){\bf A}=-i\hbar{\bgreek \nabla}+(e/c){\bf
  A}$ with ${\bf A}=(-yB_\perp,xB_\perp,2yB_\|)/2$. $H_{\rm so}$
stands for the SOC Hamiltonian, including both the Rashba\cite{rashba}
 and  IIA\cite{vervoort,vervoort2,nestoklon} terms. Then, one obtains 
 \begin{equation}
 H_{\rm
   so}=a_0(P_x\sigma_y-P_y\sigma_x)+b_0(-P_x\sigma_x+P_y\sigma_y),
 \label{eq2}
 \end{equation}
where $a_0$ and $b_0$ represent the strengths of the Rashba and IIA
terms, respectively. The Zeeman splitting is given by $H_{\rm
  Z}=\frac{1}{2}g\mu_B(B_{\perp}\sigma_z+B_{\|}\sigma_x)$
 with $g$ being the Land\'{e}
factor. $H_{\rm E}=eEx$ is the electric field term with an electric
field applied along the
$x$ direction. 
Considering that four in-plane valleys have much higher energies,
we only need to include two out-of-plane valleys in the calculation. These
two valleys lie at $\pm\langle k_0\rangle$ along the
$z$ axis with $\langle k_0\rangle=0.85(2\pi/a_{\rm
   Si})$. Here, $a_{\rm Si}=5.43~\mathring{\rm A}$ is the
 lattice constant of silicon.\cite{culcer} $H_{\rm v}$ in
 Eq.~(\ref{eq1}) describes the coupling\cite{boykin,friesen} between
 these two valleys. For convenience, one uses the subscripts ``$z$''
 and ``$\bar{z}$'' to denote the valley at
 $\langle k_0\rangle$ and the one at $-\langle k_0\rangle$,
 respectively.

To obtain the single-electron basis functions, we define
$H_0=\frac{{P_x}^2+{P_y}^2}{2m_t}+\frac{{P_z}^2}{2m_z}+V({\bf r})$. 
The $z$-component of the Hamiltonian $H_0$ can be solved
analytically with the eigenvalues being $E_{n_{z}}=\frac{{n_z}^2{\pi}^2{\hbar}^2}{8m_za^2}$.
Here $a$ represents the half-well width. The corresponding eigenfunctions are 
 \begin{eqnarray}
  \psi_{n_{z}}(z)=\left\{
     \begin{array}{ll}
      \frac{1}{\sqrt{a}}\sin[\frac{n_z\pi}{2a}(z+a)], &\mbox{$|z|\le
        a$}\\
      0, &\mbox{otherwise}
    \end{array}
    \right.
    \label{eq3}
 \end{eqnarray}
in which the index $n_{z}$ stands for the subband along the
growth direction. In our calculation, only the first subband is
included since the others have much higher energies. It is
very difficult to solve the in-plane part of $H_0$ analytically since
the lateral confinement potential of the double QD, $V_c(x,y)$, lacks the symmetry of
rotation. As the single-dot potential can be solved
analytically,\cite{shen,wang} we solve the Hamiltonian of each dot
separately instead to obtain the in-plane part of single-electron basis
functions. We define $H_{L,R}=\frac{{P_x}^2+{P_y}^2}{2m_t}+\frac{1}{2}m_t\omega_0^2(x\pm
x_0)^2+\frac{1}{2}m_t\omega_0^2y^2$. The effective diameter can then be
expressed as $d_0=\sqrt{\hbar\pi/(m_t\omega_0)}$. In the left dot,
$H_L=\frac{{P_x}^2+{P_y}^2}{2m_t}+\frac{1}{2}m_t\omega_0^2(x+x_0)^2+\frac{1}{2}m_t\omega_0^2y^2=
\frac{1}{2m_t}({p_{x^\prime}}^2+{p_y^\prime}^2)+\frac{1}{2}m_t\Omega^2({x^\prime}^2+y^2)+
w_B(x^\prime{p_y^\prime}-yp_{x^\prime})$, where $x^\prime=x+x_0$, $p_y^\prime=p_y-w_Bx_0m_t$,
$\Omega=\sqrt{{\omega_0}^2+{\omega_B}^2}$ and
$\omega_B={eB_\perp}/{(2m_t)}$. We define
$H_L^\prime=\frac{1}{2m_t}({p_{x^\prime}}^2+{p_y^\prime}^2)+\frac{1}{2}m_t\Omega^2({x^\prime}^2+y^2)$. 
One finds that the Hamiltonian $H_L$ can be solved analytically in the
system of polar coordinates while $H_L^\prime$ can be solved
analytically in the rectangular coordinate system.\cite{fock,darwin}
As it is much easier to calculate the Coulomb interaction numerically
in the rectangular coordinate system and the term
$w_B(x^\prime{p_y^\prime}-yp_{x^\prime})$ can be treated perturbatively, we solve the Schr\"odinger
equation of $H_L^\prime$ instead of $H_L$ to obtain the
single-electron basis functions. One obtains the eigenvalues\cite{fock,darwin} 
\begin{equation}
 E_{n_{x}n_{y}}=\hbar\Omega(n_{x}+n_{y}+1),
 \label{eq4}
 \end{equation}
where $n_{x,y}=0,1,2,...$ are the
orbital quantum numbers of the $x$ and $y$ direction,
respectively. The eigenfunctions are described as 
 \begin{eqnarray}
   \nonumber
   F_{n_xn_y}^L(x,y)&=&N_{n_{x}n_{y}}e^{-\alpha^2({x^\prime}^2+y^2)/2}H_{n_{x}}(\alpha
   x^\prime)H_{n_{y}}(\alpha y)\\
&&\mbox{}\times e^{iw_Bx_0m_ty/\hbar}
    \label{eq5}
 \end{eqnarray}
with
$N_{n_{x}n_{y}}=\{\alpha^2/[\pi(2^{n_{x}+n_{y}}n_{x}!n_{y}!)]\}^{1/2}$
and $\alpha=\sqrt{m_t\Omega/\hbar}$. $H_{n_{x},n_{y}}$ are the Hermite
polynomials. Thus, the eigenfunctions in different valleys can be
expressed as
$\phi_{n_{x}n_{y}n_{z}}^{z,\bar{z},L}=F_{n_{x}n_{y}}^L(x,y)\Psi_{n_{z}}^{z,\bar{z}}({\bf
  r})=F_{n_{x}n_{y}}^L(x,y)\psi_{n_{z}}(z)e^{\pm ik_{0}z}u_{z,\bar{z}}({\bf r})$ with
$u_{z,\bar{z}}({\bf r})$ representing the lattice-periodic Bloch
 functions.\cite{culcer} Then one obtains a set of single-electron basis functions
$\{\{\phi_{n_{x}n_{y}n_{z}}^{z,\bar{z},L}\},\{\phi_{n_{x}n_{y}n_{z}}^{z,\bar{z},R}\}\}$
where $\{\phi_{n_{x}n_{y}n_{z}}^{z,\bar{z},R}\}$ are the
  eigenfunctions in different valleys in the right dot and can be
  obtained by replacing $L$ and $x_0$ in the left dot by $R$ and $-x_0$. 
The orbital effect of the parallel magnetic
field is negligible due to a strong confinement along the growth direction.

In the present work, only $H_{\rm v}$ is considered to contribute to
the intervalley coupling since the overlap between the wavefunctions in
different valleys is negligibly small.\cite{culcer}
 However, there still remain some controversies over the valley
coupling nowadays.\cite{friesen,saraiva,chutia} Here, we take $\langle\Psi_{n_z}^{z,\bar{z}}|H_{\rm
   v}|\Psi_{n_z}^{\bar{z},z}\rangle=\Delta^1_{n_z}$ and $\langle\Psi_{n_z}^{z,\bar{z}}|H_{\rm
   v}|\Psi_{n_z}^{z,\bar{z}}\rangle=\Delta^0_{n_z}$ according to
 Ref.~\onlinecite{friesen}. Including this intervalley coupling, the
 single-electron eigenstates in the left dot become $\phi_{n_{x}n_{y}n_{z}}^{\pm
   L}=\frac{1}{\sqrt{2}}(\phi_{n_{x}n_{y}n_{z}}^{z,L}\pm \phi_{n_{x}n_{y}n_{z}}^{\bar{z},L})$ with
 eigenvalues
 $E_{n_{x}n_{y}n_{z}}^{\pm}=E_{n_{x}n_{y}}+E_{n_{z}}+\Delta_{n_z}^0\pm
 |\Delta_{n_z}^1|$. In these formulas,
\begin{eqnarray}
 \Delta_{n_z}^0&=&\frac{V_{\rm v}{n_z}^2{\pi}^2{\hbar}^2}{4m_za^3},\\
 \Delta_{n_z}^1&=&\frac{V_{\rm
     v}{n_z}^2{\pi}^2{\hbar}^2\cos(2k_0a)}{4m_za^3},
 \label{eq6}
 \end{eqnarray} 
with $V_{\rm v}$ representing the ratio of the valley coupling
strength to the depth of quantum well.\cite{friesen} For the case of
the right dot, one can get the corresponding single-electron
eigenvalues and eigenfunctions by replacing $L$ and $x_0$ in the left
dot by $R$ and $-x_0$. Then one obtains a new set of single-electron basis
functions $\{\{\phi_{n_{x}n_{y}n_{z}}^{\pm L}\},\{\phi_{n_{x}n_{y}n_{z}}^{\pm R}\}\}$. 
It is noted that these basis functions are nonorthogonal and
over-complete.

Then we turn to the system of two-electron double QDs, where the
total Hamiltonian is given by 
\begin{equation}
 H_{\rm tot}=(H_{\rm e}^1+H_{\rm e}^2+H_{\rm C})+H_{\rm p}+H_{\rm ep}^1+H_{\rm
   ep}^2.
 \label{eq7}
 \end{equation}
Here, the two electrons are denoted by $``1"$ and $``2"$. The
electron-electron Coulomb interaction is given by
 $H_{\rm C}=\frac{e^2}{4\pi\epsilon_0\kappa|{\bf r_1}-{\bf
     r_2}|}$ with $\kappa$ standing for the relative static dielectric
 constant. $H_{\rm p}=\sum_{{\bf q}\lambda}\hbar\omega_{{\bf
     q}\lambda}a_{{\bf q}\lambda}^+a_{{\bf q}\lambda}$ represents the
 phonon Hamiltonian with $\lambda$ and ${\bf q}$ denoting the phonon
 mode and the momentum, respectively. The electron-phonon interaction
 Hamiltonian is described by $H_{\rm ep}=\sum_{{\bf q}\lambda}M_{{\bf
     q}\lambda}(a_{{\bf q}\lambda}^++a_{-{\bf q}\lambda})e^{i{\bf
     q}\cdot{\bf r}}$ and $H_e^i$ $(i=1,2)$ is given by Eq.~(\ref{eq1}).

On the basis of the set of single-electron basis functions
$\{\{\phi_{n_{x}n_{y}n_{z}}^{\pm L}\},\{\phi_{n_{x}n_{y}n_{z}}^{\pm R}\}\}$, we construct the
two-electron basis functions in the form of either singlet or
triplet. For example, we use two single-electron spatial wavefunctions
$|n_{x1}n_{y1}n_{z1}n_{v1}p_1\rangle$ and
$|n_{x2}n_{y2}n_{z2}n_{v2}p_2\rangle$ (denoted as $|N_1\rangle$ and
$|N_2\rangle$ for short; $n_v=\pm$; $p=L/R$) to obtain the singlet
wavefunctions 
\begin{equation}
  |S^{(\Xi)}\rangle=(|\uparrow\downarrow\rangle
-|\downarrow\uparrow\rangle)\otimes
  \begin{cases}
    \frac{1}{\sqrt 2}|N_1N_2\rangle,& N_1=N_2\\
    \frac{1}{2}(|N_1N_2\rangle+|N_2N_1\rangle)
    ,& N_1\not=N_2,
  \end{cases}
  \label{eq8}
\end{equation} 
and the triplet wavefunctions for $N_1\ne N_2$
\begin{eqnarray}
 &&|T_+^{(\Xi)}\rangle=\frac{1}{\sqrt2}(|N_1N_2\rangle-|N_2N_1\rangle)\otimes|\uparrow\uparrow\rangle,\\
 &&|T_0^{(\Xi)}\rangle=\frac{1}{2}(|N_1N_2\rangle-|N_2N_1\rangle)\otimes(|\uparrow\downarrow\rangle
 +|\downarrow\uparrow\rangle),\\
 &&|T_-^{(\Xi)}\rangle=\frac{1}{\sqrt2}(|N_1N_2\rangle-|N_2N_1\rangle)\otimes|\downarrow\downarrow\rangle.
 \label{eq9_11}
\end{eqnarray}
Here, the spatial wavefunctions of the first and second electrons in
$|NN'\rangle$ are denoted as $N$ and $N'$ in sequence. Specially, we
denote $p_1=p_2=L/R$ as $(2,0)/(0,2)$ and $p_1\ne p_2$
as $(1,1)$ configuration. The superscript $(\Xi)$ denotes the valley
  configuration of each state. We define $\Xi=\pm$ for the valley indices
  of single electron states $n_{v1}=n_{v2}=\pm$, and $\Xi=m$ for
$n_{v1}\ne n_{v2}$.

Then, one can calculate the matrix elements of two-electron
Hamiltonian $H_{\rm e}^1+H_{\rm e}^2+H_{\rm C}$ in Eq.~(\ref{eq7})
and obtain the two-electron Hamiltonian matrix, where the matrix elements of the Coulomb
interaction can be expressed by
\begin{widetext}
 \begin{equation}
\langle N_1N_2|H_{\rm C}|N_1^{\prime}N_2^{\prime}\rangle=\frac{e^2}{32{\pi}^3\epsilon_0\kappa}
\sum_{\gamma_{1},\gamma_{2},\gamma_1^{\prime},\gamma_2^{\prime}=z,\bar{z}}
\eta_{n_{v1}}^{\gamma_{1}}\eta_{n_{v2}}^{\gamma_{2}}
\eta_{n_{v1}^\prime}^{\gamma_1^{\prime}}
\eta_{n_{v2}^\prime}^{\gamma_2^{\prime}}
G(\phi_{n_{x1}n_{y1}n_{z1}}^{\gamma_1,p_1},\phi_{n_{x2}n_{y2}n_{z2}}^{\gamma_2,p_2},
\phi_{n_{x1}^\prime n_{y1}^\prime n_{z1}^\prime}^{\gamma_1^\prime,p_1^\prime},
\phi_{n_{x2}^\prime n_{y2}^\prime n_{z2}^\prime}^{\gamma_2^\prime,p_2^\prime}),
  \label{eq12}
\end{equation}
where the superscripts $\gamma_i$ and $\gamma_i^\prime$ run over the two valleys,
$z$ and $\bar z$, with $\eta_\pm^z=1$ and
$\eta_+^{\bar z}=-\eta_-^{\bar z}=1$. $G$ is given by
\begin{equation}
G(\phi_{n_{x1}n_{y1}n_{z1}}^{\gamma_1,p_1}
,\phi_{n_{x2}n_{y2}n_{z2}}^{\gamma_2,p_2},
\phi_{n_{x1}^\prime n_{y1}^\prime n_{z1}^\prime}^{\gamma_1^\prime,p_1^\prime},
\phi_{n_{x2}^\prime n_{y2}^\prime
  n_{z2}^\prime}^{\gamma_2^\prime,p_2^\prime})=\int d^3{\bf k}\frac{\langle\phi_{n_{x1}n_{y1}n_{z1}}^{\gamma_1,p_1}|e^{i{\bf k}\cdot{\bf r}}|\phi_{n_{x1}^\prime n_{y1}^\prime
    n_{z1}^\prime}^{\gamma_1^\prime,p_1^\prime}\rangle\langle\phi_{n_{x2}^\prime
    n_{y2}^\prime n_{z2}^\prime}^{\gamma_2^\prime,p_2^\prime}|e^{i{\bf
      k}\cdot{\bf r}}|\phi_{n_{x2}n_{y2}n_{z2}}^{\gamma_2,p_2}\rangle^*}{k^2}
\end{equation}
\label{eq13}
\end{widetext}
As two-electron basis functions are
nonorthogonal, we also calculate the overlap between these basis
functions. One finds
  that these two-electron basis functions can be divided
  into three independent subspaces according to the valley index, i.e.,
  $\Xi=\pm$ and $m$, as there is nearly no couping between them due
  to the negligibly small intervalley Coulomb interaction\cite{li} and
  overlap between the wave functions in different
  valleys.\cite{culcer} Then one can diagonalize the
eigen equation $\tilde{H}^{(\Xi)}X=\lambda \tilde{S}^{(\Xi)}X$ in each subspace
separately and obtain corresponding two-electron energy
spectra and eigenfunctions, where $\tilde{H}^{(\Xi)}$
and $\tilde{S}^{(\Xi)}$ stand for the two-electron Hamiltonian and overlap
matrix in the subspace with the valley index $\Xi$ ($\Xi=\pm$ or
$m$) respectively.\cite{lowdin} We identify a two-electron eigenstate as
singlet (triplet) if its amplitude of the singlet (triplet) components
is larger than 50~\%. We use the similar way to identify a two-electron eigenstate as
$(2,0)$, $(0,2)$ or $(1,1)$ configuration according to the maximum amplitude.

The transition rate from the state
$|i\rangle$ to $|f\rangle$ due to the
electron-phonon scattering is calculated by the Fermi golden rule,
\begin{eqnarray}
  \Gamma_{i\rightarrow f}&=&\frac{2\pi}{\hbar}\sum_{{\bf
      q}\lambda}|M_{{\bf q}\lambda}|^2|\langle f|\chi|i\rangle|^2[\bar{n}_{{\bf
      q}\lambda}\delta(\epsilon_f-\epsilon_i-\hbar\omega_{{\bf
      q}\lambda})\nonumber\\
  &&\mbox{}+(\bar{n}_{{\bf
      q}\lambda}+1)\delta(\epsilon_f-\epsilon_i+\hbar\omega_{{\bf
      q}\lambda})],
  \label{eq14}
\end{eqnarray}
in which $\chi({\bf q},{\bf r_1},{\bf r_2})=e^{i{\bf q}\cdot{\bf
    r_1}}+e^{i{\bf q}\cdot{\bf r_2}}$ and
$\bar{n}_{{\bf q}\lambda}$ stands for the Bose distribution of phonons. In
the calculation, the temperature is fixed at 0~K and only the
second term, i.e., the phonon-emission process, occurs. One finds that
the transition between the eigenstates in different subspaces is
almost forbidden because $\langle f|\chi|i\rangle$ in Eq.~(\ref{eq14}) is
strongly suppressed due to a large intervalley wave vector $\langle
2k_0\rangle$, similar to the suppressions of the intervalley
Coulomb interaction and overlap between the wave functions in
different valleys mentioned above.

\section{NUMERICAL RESULTS}
From the Fermi golden rule [Eq.~(\ref{eq14})], one finds that the phonon energy is just the energy
difference between the initial and final electron states. The energy difference studied here is much
smaller than the energies of the intervalley acoustic phonon and the
optical phonon.\cite{pop} Besides, the piezoelectric interaction is absent in
silicon,\cite{li} therefore one only needs to take into account the intravalley
electron-acoustic phonon scattering due to the deformation
potential. In this work, both the TA and LA phonons are included. The corresponding matrix
elements read $M_{\beta,{\rm intra},{\bf
    Q}}^2={\hbar D_{\beta}^2Q^2}/{(2d\Omega_{\beta,{\rm intra},{\bf
      Q}})}$ with $\beta$=LA/TA representing the LA/TA phonon
mode. The deformation potentials for the LA and TA phonons are $D_{{\rm LA}}=3.93$~eV
and $D_{{\rm TA}}=2.48$~eV, respectively.\cite{pop} The
mass density of silicon $d=2.33$~g/cm$^3$.\cite{sonder}
The phonon energy $\Omega_{\beta,{\rm intra},{\bf Q}}=v_{\beta}Q$ with
sound velocities $v_{\rm LA}=9.01\times 10^5$~cm/s and $v_{\rm
  TA}=5.23\times 10^5$~cm/s.\cite{pop} 
The effective mass $m_t=0.19m_0$ and $m_z=0.98m_0$ with $m_0$
being the free electron mass.\cite{dexter} The Land\'{e} factor
$g=2$,\cite{graeff} the ratio of the valley coupling strength to the depth of
quantum well $V_{\rm v}=7.2\times 10^{-11}$~m,\cite{friesen} and the relative
static dielectric constant $\kappa=11.9$.\cite{sze} In the
  previous work on silicon double QDs,\cite{li,culcer,das} only the 
    lowest few basis functions were included in the calculation. We
  find that these basis functions are enough for the convergence of
  the energy spectra, but inadequate in obtaining the correct transition
rates. The energy spectra calculated with the lowest few basis
  functions and the convergent ones differ by about $0.1$~\%, but the
  transition rates calculated with the lowest few basis functions
differ by about four orders of magnitude from the convergent
ones. Therefore, in our calculation, we employ the
exact-diagonalization method with the lowest 1050 singlet and 3060
triplet basis functions to ensure the convergence of the eigenstates and
the transition rates. It is noted that all the basis functions
chosen are in the subspace with the valley index $``-"$ since
 we focus on the case
of a large valley splitting where a large effective diameter is taken to
    make sure that the lowest singlet and triplet states under
  investigation are in the subspace with the valley index
  $``-"$. This choice does not lead to the loss of
  generality as states with different valley indices are nearly decoupled as
  pointed out above.
It is also noted that the Coulomb interaction was treated as
a Hubbard parameter $u$ in the previous works on silicon double
QDs,\cite{li,culcer,das} but in our work, it is explicitly
calculated.  One can obtain $u$ from our
calculation, e.g., $u$ in the model by Culcer
{\em et al.}\cite{culcer} is determined to be about $23$~meV.

\subsection{PERPENDICULAR MAGNETIC-FIELD DEPENDENCE}
We first investigate the case of a perpendicular magnetic field. We
take 32 monoatomic layers of silicon along the growth direction of the
quantum well, corresponding to the well width
$2a=4.344$~nm. According to Eq.~(\ref{eq6}), a large valley splitting
$2|\Delta_{n_z}^1|=0.83$~meV is obtained. Then we choose the
  effective diameter $d_0=30$~nm.
With an electric field 30~kV/cm along the growth direction, 
one obtains the strength of the Rashba SOC
induced by this electric field $a_0=-6.06$~m/s and that of the IIA term
$b_0=-30.31$~m/s.\cite{nestoklon}

\begin{figure}[bth]
\includegraphics[width=7.5cm]{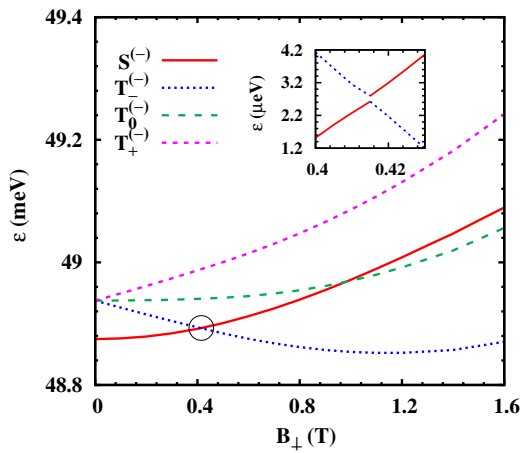}
\caption{(Color online) The lowest four energy levels {\em vs}.
  perpendicular magnetic field $B_{\perp}$ in double QDs. The anticrossing
  point between $|S^{(-)}\rangle$ and $|T_-^{(-)}\rangle$ is shown and
the range near this point is enlarged in the inset (the energies are
substracted by $48.89$~meV). In the
  calculation, the interdot distance $2x_0=20$~nm.}  
\label{fig2}
\end{figure}

\begin{figure}[bth]
\includegraphics[width=8.5cm]{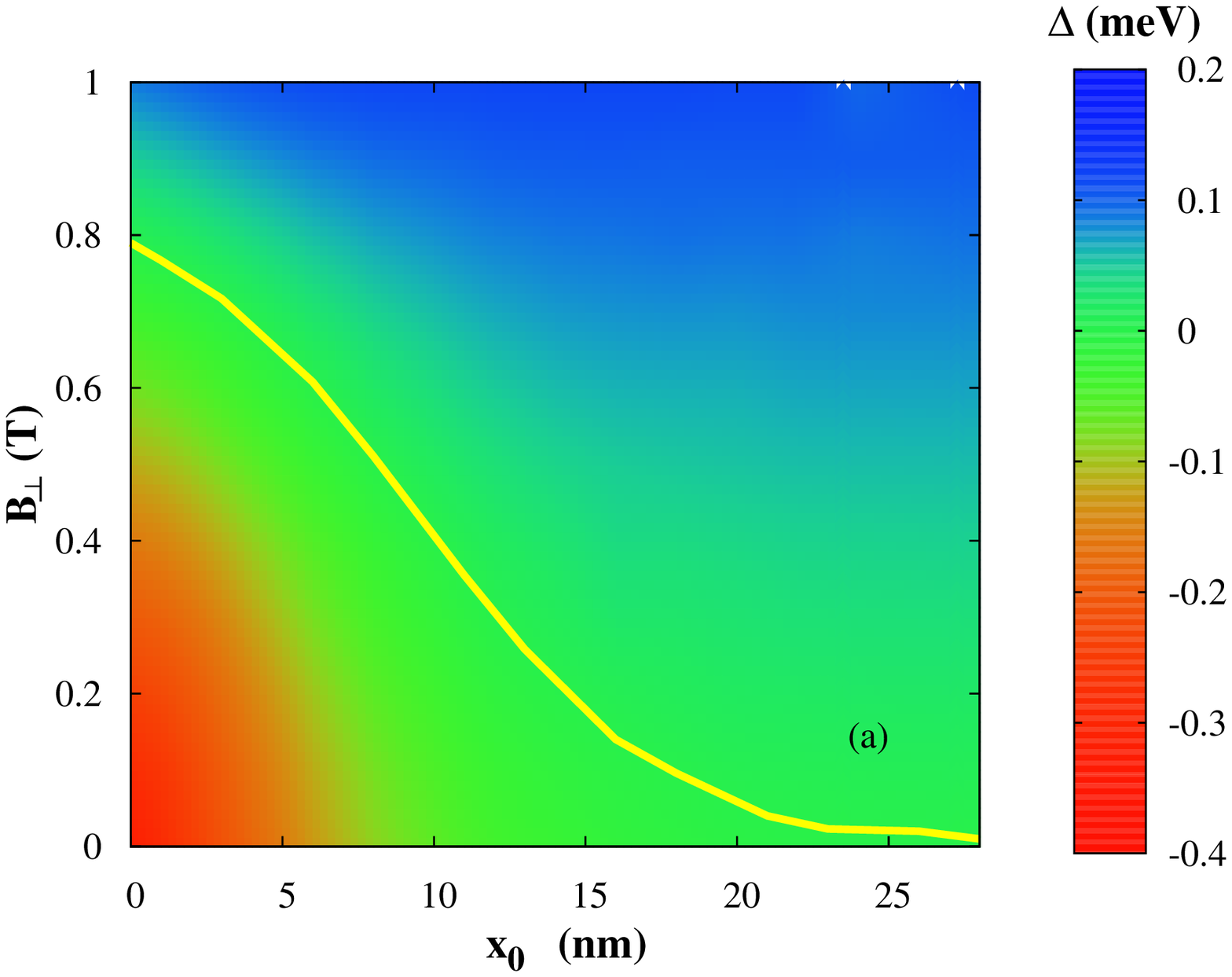}
\includegraphics[width=7.5cm]{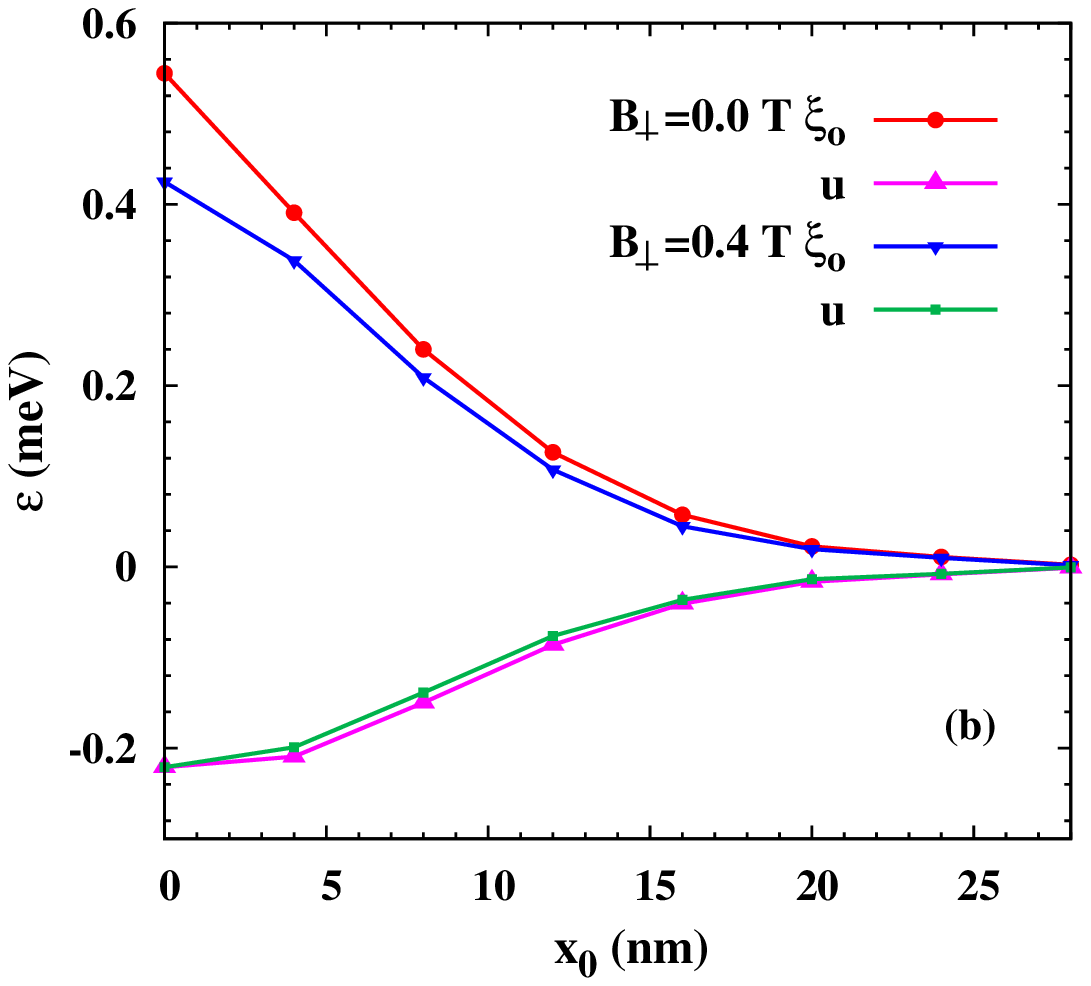}
\caption{(Color online) (a) The energy difference between the lowest singlet
  $|S^{(-)}\rangle$ and the lowest triplet $|T_-^{(-)}\rangle$ {\em vs}.
  perpendicular magnetic field $B_{\perp}$ and half of the interdot distance $x_0$ in double QDs.
  The yellow solid cure is the position of the anticrossing between
  $|S^{(-)}\rangle$ and $|T_-^{(-)}\rangle$. (b) the orbital-energy
  difference between $|T_-^{(-)}\rangle$ and $|S^{(-)}\rangle$ $\xi_o$ and the energy difference of the Coulomb
  interaction $u$ {\em vs}. half of the interdot distance $x_0$ at
 zero magnetic field and a magnetic field $B_{\perp}=0.4$~T.}   
\label{fig3}
\end{figure}

\begin{figure}[bth]
\includegraphics[width=7.5cm]{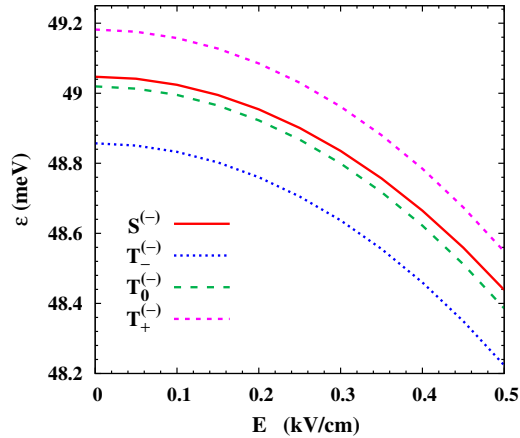}
\caption{(Color online) The lowest four energy levels {\em vs}. electric
  field along the $x$ direction. In the calculation, the interdot
  distance $2x_0=20$~nm and the magnetic
  field $B_{\perp}=1.4$~T.}  
\label{fig4}
\end{figure}

With the interdot distance $2x_0=20$~nm, 
 the lowest four levels are plotted in
Fig.~\ref{fig2} as function of the perpendicular magnetic field,
denoted as $|S^{(-)}\rangle$,
$|T_+^{(-)}\rangle$ (spin up), $|T_0^{(-)}\rangle$ (zero spin) and $|T_-^{(-)}\rangle$ (spin down)
according to their major components. The energies of three triplet states are separated by
the Zeeman splitting. From the figure, one finds that the singlet state $|S^{(-)}\rangle$
intersects with the triplet states $|T_-^{(-)}\rangle$ and $|T_0^{(-)}\rangle$ in
sequence with the increase of the magnetic field. The intersecting point between
$|S^{(-)}\rangle$ and $|T_-^{(-)}\rangle$ ($B_{\perp}\sim 0.415$~T)
is an anticrossing point where there exists a small energy gap ($\sim
0.17$~$\mu {\rm eV}$) shown in the inset due to the Rashba
SOC\cite{rashba} and the IIA term.\cite{nestoklon} The 
intersecting point between $|S^{(-)}\rangle$ and $|T_0^{(-)}\rangle$
($B_{\perp}\sim 0.98$~T) is simply a crossing point.

The anticrossing between $|S^{(-)}\rangle$ and $|T_-^{(-)}\rangle$ can also be tuned
by varying the interdot distance. We plot the energy difference between
$|S^{(-)}\rangle$ and $|T_-^{(-)}\rangle$ as function of the magnetic
field and the interdot distance in Fig.~\ref{fig3}(a). In this figure, we
also show the position of the anticrossing between $|T_-^{(-)}\rangle$ and 
$|S^{(-)}\rangle$. It is seen that the magnetic field where the
anticrossing occurs decreases with the increase of the interdot
distance. This can be understood from the energy difference between
$|T_-^{(-)}\rangle$ and $|S^{(-)}\rangle$: $\Delta(B_{\perp},x_0)=\xi_o+u-g\mu_BB_{\perp}$, where
$\xi_o$ is the orbital-energy difference and $u$ comes from the contribution of
the Coulomb interaction. By solving the equation
$\Delta(B_{\perp},x_0)=0$, one obtains the magnetic field $B_{\perp}^c$
corresponding to the anticrossing point. To facilitate understanding
of the dependence of $B_{\perp}^c$ on the interdot distance, we plot the
interdot-distance dependence of $\xi_o$ and $u$ at zero and a specific
magnetic fields in Fig.~\ref{fig3}(b). From this figure, one finds that the
orbital-energy difference $\xi_o$ decreases with increasing the
interdot distance while $u$, which is
insensitive to the magnetic field under investigation, shows an opposite behavior. It is noted
that the increase of $u$ is much smaller than the decrease of
the orbital-energy difference $\xi_o$. Therefore, with the increase of the interdot distance,
the net contribution of $\xi_o+u$ decreases and correspondingly
the magnetic field where the anticrossing occurs, i.e., $B_{\perp}^c$,
decreases too.

In addition, the electric field can also effectively affect the energy levels. We apply an electric field
along the $x$ direction and plot the lowest four levels in
Fig.~\ref{fig4}. One notices that the energy
levels are weakly dependent on the electric field in the small electric
field regime but show a rapid decrease when the electric field becomes
strong. This electric field dependence agrees with that reported by
Culcer {\em et al.}\cite{culcer} qualitatively. Moreover, one also finds
that the energy differences among these levels are almost independent
of the electric field. These behaviors can be understood as
follows. With the increase of the electric field, the configuration of 
these four states gradually changes from $(1,1)$ to $(2,0)$ according
to their major components. In $(1,1)$ configuration, i.e., in the small
electric field regime, the electric field suppresses the single-electron
energy in the left dot while raises it in the
right dot. Therefore, the net contribution of the electric field is small
and changes slowly with increasing electric field. However, when
the electric field becomes strong, i.e., the states are in $(2,0)$
configuration, the energy of each electron decreases while that of the
two-electron Coulomb interaction increases with increasing
electric field. It is noted that the increase of the
energy of the Coulomb interaction is much smaller than the decrease of the energies
induced by the electric field. As a result, the energy levels show a
rapid decrease. Besides, we find that these four states always keep
the same configurations regardless of the strength of the
electric field. Therefore, the electric field has the same effect on
these states, leading to the energy differences among these levels
insensitive to the electric field.

\begin{figure}[bth]
\includegraphics[width=7.5cm]{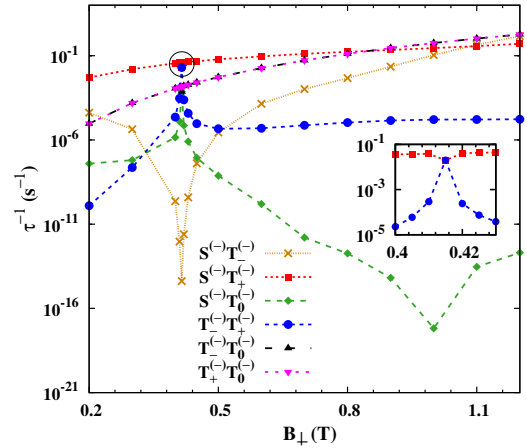}
\caption{(Color online) Transition rates {\em vs}. perpendicular
  magnetic field. The inset zooms the range near the
  anticrossing point of the channels $S^{(-)}T_+^{(-)}$ and
  $T_-^{(-)}T_+^{(-)}$. In the calculation, the interdot distance $2x_0=20$~nm.}  
\label{fig5}
\end{figure}

\begin{figure}[bth]
\includegraphics[width=7.5cm]{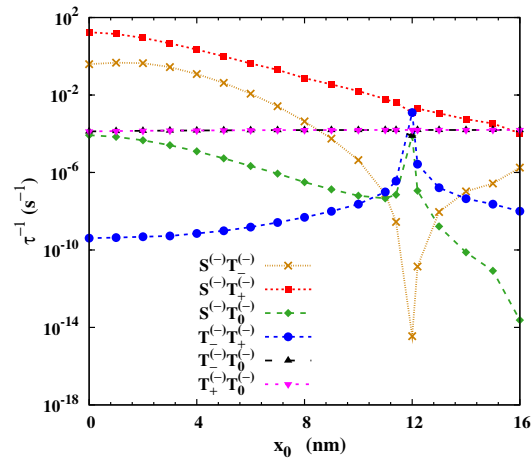}
\caption{(Color online) Transition rates {\em vs}. half of the interdot distance. In the
  calculation, the magnetic field $B_{\perp}=0.3$~T.}  
\label{fig6}
\end{figure}

We then calculate the ST relaxation rates together with the transition rates of the
channels between two triplet states. From Fig.~\ref{fig5}, one finds that the transition rates can be markedly 
modulated by the magnetic field. In the vicinity of the anticrossing point
($B_{\perp}\sim 0.415$~T), the transition rates show intriguing
features. The rate between $|S^{(-)}\rangle$ and $|T_-^{(-)}\rangle$
shows a sharp decrease due to small phonon energy, which has been
addressed in our previous investigations on GaAs and Si single
QDs.\cite{shen,wang} The transition rates of other 
channels except the one between $|T_+^{(-)}\rangle$ and
$|T_0^{(-)}\rangle$ present either a peak or a valley due to the large spin
mixing between the singlet $|S^{(-)}\rangle$ and the triplet
$|T_-^{(-)}\rangle$, similar to the behavior we have investigated in
single QDs.\cite{wang} Specifically, the transition rate between
$|S^{(-)}\rangle$ and $|T_+^{(-)}\rangle$ and the one between
$|T_-^{(-)}\rangle$ and $|T_0^{(-)}\rangle$ present a minimum while
the transition rate between $|S^{(-)}\rangle$ and $|T_0^{(-)}\rangle$
and the one between $|T_-^{(-)}\rangle$ and $|T_+^{(-)}\rangle$ show a
maximum. Far away from the anticrossing point, the 
variation of the transition rates can be well understood from the change
of the phonon energy.\cite{climente,shen,wang}

We also investigate the influence of the interdot distance on
the ST relaxation rates together with the transition rates of the
channels between two triplet states at $B_{\perp}=0.3$~T. The results are shown in
Fig.~\ref{fig6}. We also find an anticrossing point between
$|S^{(-)}\rangle$ and $|T_-^{(-)}\rangle$ at $x_0\sim 12$~nm. In the vicinity
of this point, the behavior of the transition rates is similar to what
we have obtained above by sweeping the magnetic field. The
transition rate between $|S^{(-)}\rangle$ and $|T_-^{(-)}\rangle$ is strongly
suppressed and other transition rates relevant to these two states also show a
rapid increase or decrease. Therefore, one can tune the ST relaxation by varying the
interdot distance which can be controlled electrically in the experiment.

Moreover, the electric-field dependence of the transition rates is
also studied. We find that the transition rates are almost
independent of the electric field (not shown). This behavior can 
be understood from Fig.~\ref{fig4} where the energy
differences among the lowest four levels are almost independent of the
electric field.\cite{shen,wang} This property is of great importance
in the spin manipulation, since the lifetime remains almost identical
during the change of the qubit configuration from $(1,1)$ to $(2,0)$
by electric field.

\begin{figure}[bth]
\includegraphics[width=7.5cm]{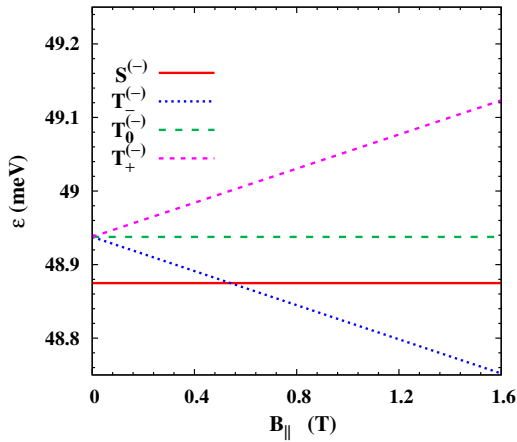}
\caption{(Color online) The lowest four energy levels {\em vs}.
  parallel magnetic field $B_{\|}$ in double QDs. In the
  calculation, the interdot distance $2x_0=20$~nm.}  
\label{fig7}
\end{figure}

\begin{figure}[bth]
\includegraphics[width=7.5cm]{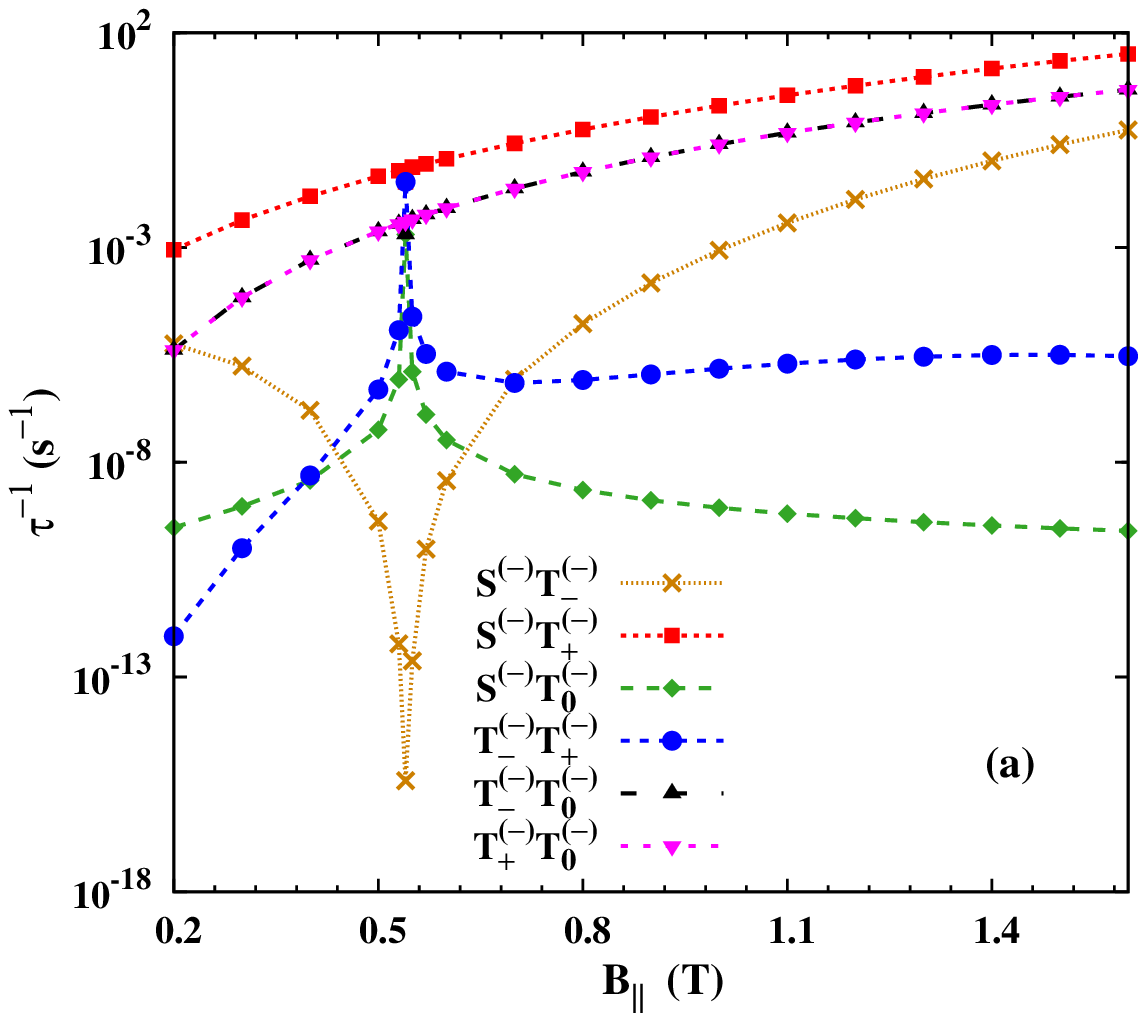}
\includegraphics[width=7.5cm]{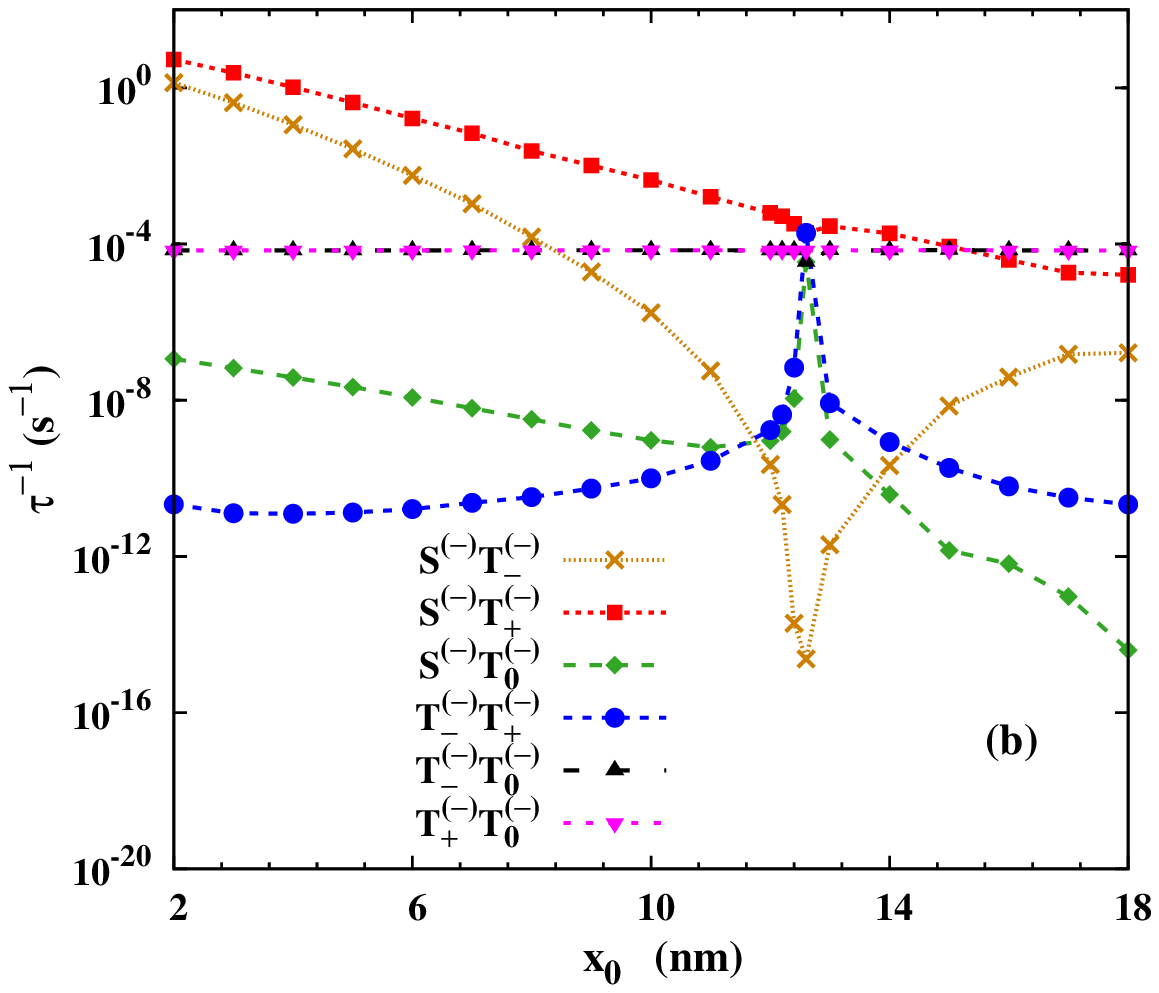}
\caption{(Color online) (a) Transition rates {\em vs}. parallel
  magnetic field $B_{\|}$ with the interdot distance $2x_0=20$~nm. (b)
  Transition rates {\em vs}. half of the interdot distance with the magnetic field $B_{\|}=0.3$~T.}  
\label{fig8}
\end{figure}

\subsection{PARALLEL MAGNETIC-FIELD DEPENDENCE}
We also investigate the case with a parallel
magnetic field along the $x$ direction. The well width, effective
diameter and strengths of both the Rashba SOC and the IIA term are
chosen to be the same as the perpendicular
magnetic-field case. We investigate the magnetic-field
dependence of the energy spectrum in the absence of the applied electric 
field  with the interdot distance
$2x_0=20$~nm. The lowest four levels are plotted in
Fig.~\ref{fig7}. We denote these states as $|S^{(-)}\rangle$,
$|T_+^{(-)}\rangle$, $|T_0^{(-)}\rangle$ and $|T_-^{(-)}\rangle$
according to their major components. From the figure, one finds that
$|S^{(-)}\rangle$ and $|T_0^{(-)}\rangle$ are almost independent of
the magnetic field while $|T_+^{(-)}\rangle$ and $|T_-^{(-)}\rangle$
show a linear dependence. This is because of the absence of the
orbital effect of the parallel magnetic field thanks to a
strong confinement along the growth direction. Therefore, the
magnetic-field dependence is involved only through the Zeeman
splitting. For $|S^{(-)}\rangle$ and $|T_0^{(-)}\rangle$, the
$x$-components of the total spin are almost zero, leading to negligibly
small Zeeman splitting. For $|T_+^{(-)}\rangle$ and
$|T_-^{(-)}\rangle$, the $x$-components of the total spin are nearly $\pm 1$, 
which indicate that these two levels change linearly with the magnetic
field. Besides, we also find an anticrossing point between
$|S^{(-)}\rangle$ and $|T_-^{(-)}\rangle$ at $B_{\|}\sim 0.54$~T due to the SOCs.

The influence of the magnetic field and interdot
distance on the ST relaxation rates is investigated. From Fig.~\ref{fig8}, one finds that the
behavior of the transition rates is quite similar to what obtained in
the case of the perpendicular magnetic field. Here, one also observes the anticrossing point
between $|S^{(-)}\rangle$ and $|T_-^{(-)}\rangle$ by sweeping the
parallel magnetic field and/or interdot distance. In the vicinity of the
anticrossing point, the transition rates relevant to these two states
also show a rapid increase or decrease.

\section{SUMMARY}
In summary, we have investigated the ST relaxation in two-electron
silicon double QDs with magnetic fields in both the Faraday and
Voigt configurations.
 The electron-electron Coulomb interaction and the mutivalley
effect are explicitly included. 
A large number of basis functions are utilized to
converge the eigenstates and the transition rates. 
We find that the external magnetic
field and the interdot distance have strong influence on the lowest
four energy levels and consequently the transition rates 
can be effectively modulated by the external magnetic field and
the interdot distance. Moreover, from the magnetic-field
and interdot-distance dependences of the energy spectrum, we observe ST
anticrossing points. In the vicinity of the anticrossing point, a
small energy gap exists between the singlet and one of the triplet
states due to the SOCs. The transition rates of the channels relevant
to these two states show either a peak or a valley. 
Furthermore, we also
study the effect of the electric field on the energy spectra and the
transition rates. We find that the configuration of the 
lowest four levels change from $(1,1)$ to $(2,0)$ 
with the increase of the electric field. 
Differing from the magnetic-field and interdot-distance dependences,
the transition rates are nearly independent of the electric field.
This is of great importance in the spin manipulation since the lifetime
remains almost unchanged during the manipulation of qubit configuration.

\begin{acknowledgments}
This work was supported by the Natural Science Foundation of China
under Grant No.\ 10725417. One of the authors (L.W.)
would like to thank Y. Yin, P. Zhang and K. Shen for valuable discussions.
\end{acknowledgments}

\end{document}